\documentstyle{pasj00}
\SetRunningHead{A. \textsc{Imada}}{Photmetory of NSV 4838}
\Received{$\langle$reception date$\rangle$}
\Accepted{$\langle$acception date$\rangle$}
\Published{$\langle$publication date$\rangle$}
\SetRunningHead{A. \textsc{Imada} et al.}{Usage of \texttt{pasj00.cls}}

\begin{document}

\title{CCD Photometry of a Newly Confirmed SU UMa-Type Dwarf Nova, NSV 4838}

\author{Akira \textsc{Imada}, Tatsuki \textsc{Yasuda}, Toshihiro
\textsc{Omodaka}, Shota \textsc{Oizumi}, Hiroyuki \textsc{Yamamoto}, Shunsuke
\textsc{Tanada}, Yoshihiro \textsc{Arao}, Kie \textsc{Kodama}, Miho
\textsc{Suzuki}, Takeshi \textsc{Matsuo}}
\affil{Faculty of Science, Kagoshima University,1-21-30
Korimoto, Kagoshima-shi, Kagoshima 890-0065}
\email{yasuda@astro.sci.kagoshima-u.ac.jp}

\author{Hiroyuki \textsc{Maehara}, Taichi \textsc{Kato}, Kei
\textsc{Sugiyasu}, Yuuki \textsc{Moritani}, Masanao \textsc{Sumiyoshi}}
\affil{Department of Astronomy, Faculty of Science, Kyoto
University, Sakyo-ku, Kyoto 606-8502}

\author{Kazuhiro \textsc{Nakajima}}
\affil{VSOLJ, 124 Isatotyo Teradani, Kumano, Mie 519-4673}

\author{Johen \textsc{Pietz}}
\affil{Rostocker Str. 62, 50374 Erftstadt, Germany}

\author{Kenshi \textsc{Yanagisawa}}
\affil{Okayama Astrophysical Observatory, National Astronomical
Observatory of Japan, Asakuchi, Okayama 719-0232}

\and

\author{Daisaku \textsc{Nogami}}
\affil{Hida Observatory, Kyoto University, Kamitakara, Gifu 506-1314}


\maketitle

\begin{abstract}
We present time-resolved CCD photometry of a dwarf nova NSV 4838 (UMa
 8, SDSS J102320.27+440509.8) during the 2005 June and 2007 February
 outburst. Both light curves showed superhumps with a mean period of
 0.0699(1) days for the 2005 outburst and 0.069824(83) days for the 2007
 outburst, respectively. Using its known orbital period of 0.0678 days,
 we estimated the mass ratio of the system to be $q$=0.13 based on an empirical relation. Although the majority of SU UMa-type dwarf novae having similar superhump periods show negative period derivatives, we found that the superhump period increased at $\dot{P}$/$P_{\rm sh}$=+7(+3, -4)$\times$10$^{-5}$ during the 2007 superoutburst. We also investigated long-term light curves of NSV 4838, from which we derived 340 days as a supercycle of this system.
\end{abstract}

\section{Introduction}

Dwarf novae are a group of cataclysmic variable stars that consist of a
white dwarf (primary) and a late-type star (secondary) (for a review,
see e.g., \cite{war95book}; \cite{osa96review}; \cite{hel01book};
\cite{osa05review}; \cite{smi07review}). The orbital periods of the
majority of dwarf novae are below 9 hours \citep{gae05review}. The
secondary star fills its Roche lobe, transferring gas into the primary
Roche lobe via the inner Lagrangian point (L1). This process results in
formation of an accretion disk around the primary. The accretion
disk plays the main role in dwarf nova outbursts which are driven by
the thermal limit-cycle instability (\cite{hos79DI}; \cite{mei81hlcma};
\cite{can88outburst}). 

SU UMa-type dwarf novae are one subclass of dwarf novae, whose orbital
periods are shorter than 2 hours except for a few systems (for a review,
see e.g., \cite{osa89suuma}). These systems exhibit two types of
outbursts. One is normal outburst, whose duration is typically a few
days. The other is superoutburst, whose duration is longer than 10 days
and their maximum magnitude is ${\sim}$0.5 mag brighter than that of normal
outburst. During the superoutburst, the light curve shows tooth-like
modulations called superhump with an amplitude of ${\sim}$0.2 mag. The
period of the superhump ($P_{\rm sh}$) is a few percent longer than
those of the orbital period ($P_{\rm sh}$). This slightly longer
periodicity and the modulations are attributed to phase-dependent
removal of the angular momentum in a tidally deformed eccentric disk
(\cite{whi88tidal}; \cite{hir90SHexcess}; \cite{mur98SH};
\cite{smi07SH}).

NSV 4838 (UMa 8 \citep{dow01cvcat}, SDSS J102320.27+440509.8 \citep{szk05CVSDSS}) was listed by \citet{1977A&AS...28..123B} as a class ${\rm I}$ (UV-richest) object. \citet{DownesCVatlas2} introduced this object as a
candidate dwarf nova or a nova-like star with its magnitude ranging
14.5p-16.5p. An optical spectrum was published by \citet{szk05CVSDSS},
in which Balmer emission lines are superimposed on a
blue continuum. \citet{szk05CVSDSS} also gave the SDSS quiescent magnitudes of the
object as $u$=18.513, $g$=18.830, $r$=18.688, $i$=18.570, and
$z$=18.514. Radial velocity studies were implemented by Thorstensen et
al. (unpublished), who determined $P_{\rm orb}$=0.0678 days (see also,
\cite{rit03cat}). No 2MASS counterpart is detected, but recent infrared
photometry using the OAO/ISLE \citep{yan06ISLE} yields $H$=17.51(0.09)
and $K_{\rm s}$=16.81(0.08), respectively. The SU UMa nature of the
object was initially confirmed by T. Vanmunster by detection of superhumps
during the 2005 June superoutburst. Here we report on photometric
studies of the 2005 June and 2007 February superoutbursts, as well as
the long-term behavior of the object.

\section{Observations}

Time-resolved CCD photometry was performed from 2005 June 6 to June 14,
and from 2007 February 7 to 28 at 5 sites. The log of observations is
summarized in table \ref{table1}. Detailed information on sites is given
in table \ref{table2}. The total data points amounted to 7428. The
exposure times were 30-180 seconds without filter. All CCD systems are
close to the Kron-Cousins $R_{\mathrm{C}}$ band. After debiasing and
flat-fielding, we performed aperture photometry using IRAF\footnote{IRAF
(Image Reduction and Analysis Facility) is distributed by US National
Optical Observatories, which is operated by the Association of
Universities, for Research in Astromy, Inc., under cooperative agreement
with National Science Foundation.} apphot for data of KU and
Mhh, AIP4WIN \citep{aip4win} for data of JP, and
FitsPhot4.1\footnote{http://www.geocities.jp/nagai\_kazuo/dload-1.html} 
for data of Njh, respectively. The Kyoto team (Kyoto) used a Java-based
point spread function
(PSF) photometry package developed by one of the authors (TK). The
differential magnitude among each site was adjusted to that of the
KU site, where we chose the comparison star NOMAD 1340-0218270,
RA:10:23:33.34, Dec:+44:03:26.0 $V$ = 15.20, $R$=15.16). The constancy of the star is checked by nearby stars in the same frame. A heliocentric correction
was made for each dataset before the following analyses.

  \begin{table*}
  \begin{center}
  \caption{Log of observation.}
  \label{table1}
   \begin{tabular}{cccccc}
   \hline\hline   
   Date & HJD(start)$^*$ & HJD(end)$^*$ & N$^{\dagger}$ & Exp$^{\ddagger}$ & ID$^{\S}$ \\
   \hline
    2005 Jun 6  & 53528.3834 & 53528.4586 & 46 & 45-60 & JP \\
    2005 Jun 7  & 53529.3923 & 53529.5135 & 96 & 45-60 & JP \\
    2005 Jun 8  & 53530.3811 & 53530.5200 & 107 & 45-60 & JP \\
    2005 Jun 9  & 53531.3804 & 53531.5176 & 111 & 45-60 & JP \\
    2005 Jun 13 & 53535.3978 & 53535.4050 & 6 & 60 & JP \\
    2005 Jun 14 & 53536.4019 & - & 1 & 60 & JP \\
    2007 Feb 7  & 54139.2154 & 54139.3614 &  70 & 180 & KU \\
                & 54139.2237 & 54139.3387 & 210 &  30 & Njh\\
    2007 Feb 9  & 54141.2600 & 54141.2859 &  19 & 30 & Njh \\ 
   2007 Feb 10  & 54142.2642 & 54142.3415 &  35 & 180 & KU \\
   2007 Feb 11  & 54142.9669 & 54143.1278 &  76 & 180 & KU \\
                & 54143.0996 & 54143.3036 & 912 &  30 & Mhh\\
                & 54143.1802 & 54143.3029 & 224 &  30 & Njh\\
   2007 Feb 12  & 54143.9454 & 54144.3882 & 189 & 180 & KU \\
                & 54144.1465 & 54144.3098 & 844 &  30 & Mhh\\
                & 54144.1493 & 54144.3065 & 288 &  30 & Njh\\
   2007 Feb 14  & 54146.1660 & 54146.3038 & 139 &  30 & Njh\\
                & 54146.2009 & 54146.3616 &  63 & 180 & KU \\
   2007 Feb 15  & 54147.1399 & 54147.3053 & 850 &  30 & Mhh\\
   2007 Feb 16  & 54148.0700 & 54148.2768 & 445 &  30 & Kyoto\\
                & 54148.1642 & 54148.3079 & 748 &  30 & Mhh\\
   2007 Feb 18  & 54150.0437 & 54150.2397 & 1028 &  30 & Mhh\\
                & 54150.2121 & 54150.3842 &  79 & 180 & KU \\
   2007 Feb 19  & 54150.9889 & 54151.1981 & 457 &  30 & Kyoto\\
                & 54151.2283 & 54151.3833 &  74 & 180 & KU \\
   2007 Feb 20  & 54152.1114 & 54152.3373 & 100 & 180 & KU \\
   2007 Feb 21  & 54153.2148 & 54153.2973 & 40  & 180 & KU \\
   2007 Feb 23  & 54155.0150 & 54155.1822 & 80  & 180 & KU \\
   2007 Feb 25  & 54157.3127 & 54157.3804 & 33  & 180 & KU \\
   2007 Feb 26  & 54158.2717 & 54158.3311 & 28  & 180 & KU \\
   2007 Feb 28  & 54160.1487 & 54160.2101 & 30  & 180 & KU \\
   \hline
   \multicolumn{6}{l}{$^*$ HJD - 2450000. $^\dagger$ Number of frames.} \\
   \multicolumn{6}{l}{$^\ddagger$ Exposure time in unit of
   second. $^{\S}$Observers' ID, see table \ref{table2}.} \\
   \end{tabular}
  \end{center}
  \end{table*}

 \begin{table}
 \begin{center}
 \caption{List of observers.}
 \label{table2}
  \begin{tabular}{cccc}
   \hline\hline
ID & Observer & Site & Telescope \\
\hline
JP & J. Pietz & Erftstadt, Germany & 20$cm$ \\
KU  & T. Yasuda+$^*$ & Kagoshima, Japan &   100$cm$  \\
Kyoto  & A. Imada+$^{\dagger}$ & Kyoto, Japan   & 40$cm$ \\
Mhh  & H. Maehara & Saitama, Japan & 25$cm$ \\
Njh  & K. Nakajima & Mie, Japan & 25$cm$  \\
\hline
\multicolumn{4}{l}{$^*$T. Yasuda, S. Oizumi, H. Yamamoto, S. Tanada,} \\
\multicolumn{4}{l}{Y. Arao, K. Kodama, M. Suzuki, and T. Matsuo.} \\
\multicolumn{4}{l}{$^{\dagger}$A. Imada, T. Kato, and K. Sugiyasu.} \\
\end{tabular}
 \end{center}
 \end{table}

\section{Results}

\subsection{2007 February superoutburst}

\subsubsection{light curve}

\begin{figure}
   \begin{center}
\resizebox{80mm}{!}{\includegraphics{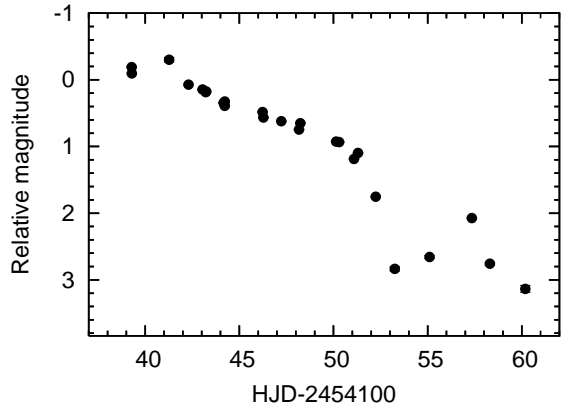}}
   \end{center}
   \caption{Averaged light curve of the 2007 February superoutburst. The
 horizontal and vertical axes indicate the HJD and the relative
 magnitude, respectively. The filled circles represent the averaged
 magnitudes for each run. The size of the standard error bar is
 comparable to that of the datapoints. The magnitude obtained on HJD
 2454141 deviates slightly from the general trend, possibly due to
 sparse data and an unfavorable weather. A hint of a rebrightening was
 seen on HJD 2454157. The magnitude of the comparison star is $V$ =
 15.20.}
\label{label1}
\end{figure}

The light curve obtained during the 2007 February superoutburst is shown in
Figure \ref{label1}. The plateau phase lasted at least 12
days\footnote{The last observation before this outburst was
performed at JD 2454125.8414, when the star was fainter than 16.2 in visual magnitude. Thus we were unable to constrain the duration of the superoutburst.} and the variable almost linearly faded at a rate of 0.10(1) mag d$^{\mathrm{-1}}$. After the plateau stage, NSV 4838 entered the rapid decline phase at a rate of 0.86(1) mag d$^{\mathrm{-1}}$. These values were typical for the superoutburst of SU UMa-type dwarf novae. Although the data were noisy during the post outburst stage, a hint of a rebrightening was seen on HJD 2454157.

\subsubsection{superhump}

\begin{figure}
   \begin{center}
      \FigureFile(80mm,50mm){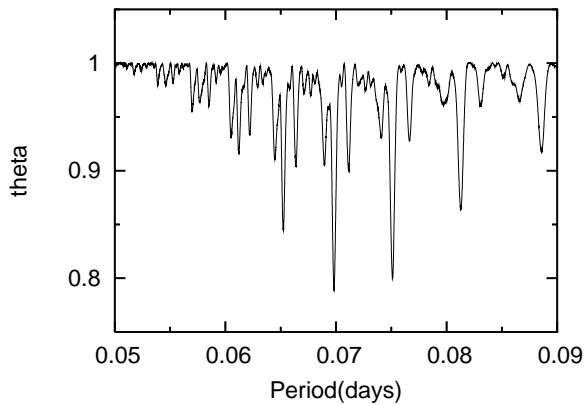}
   \end{center}
   \caption{The theta diagram of the PDM analysis. This  provides the
 best estimated period of 0.069824 d.}
\label{label2}
\end{figure}

\begin{figure}
 \begin{center}
  \resizebox{80mm}{!}{\includegraphics{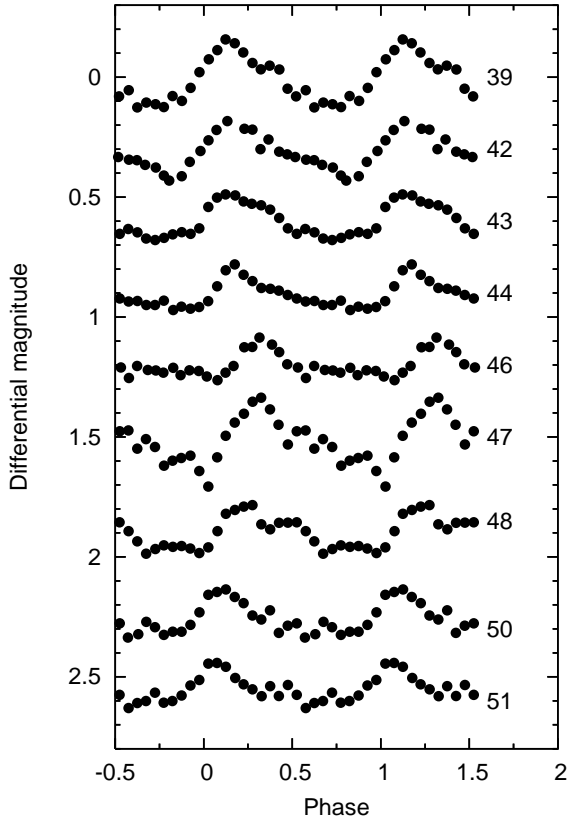}}
 \end{center}
 \caption{Daily averaged light curves of the 2007 February
 superoutburst. The numbers in this figure indicate
 HJD-2454100.}\label{label3}
\end{figure}

\begin{figure}
\begin{center}
\resizebox{80mm}{!}{\includegraphics{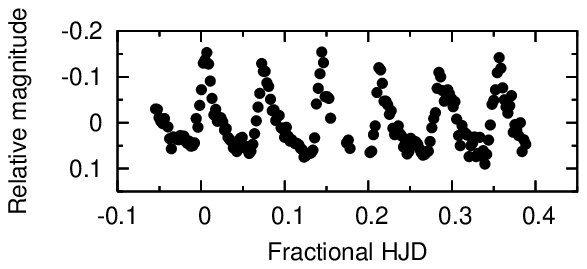}}
\end{center}
\caption{Enlarged light curve obtained by KU on 2007 February 12. A
 characteristic feature of superhumps is visible. The absence of an
 eclipse means a low-to-mid inclination of NSV 4838.}
\label{lc_rep07}
\end{figure}

In order to estimate the mean superhump period during the plateau stage,
we performed the phase dispersion minimization method (PDM, \cite{ste78pdm})
using 6750 points between 2007 February 7 and 19. Figure \ref{label2}
shows the resultant theta diagram, indicating that the best estimated
period is 0.069824(83) days. The error of the period was estimated using
the Lafler-Kinman class of methods \citep{fer89error}. Daily averaged
light curves folded by this period are displayed in figure
\ref{label3}. Also shown is a representative light curve on 2007 February
12 (HJD 2454143.9454-2454144.3882) obtained at KU (figure
\ref{lc_rep07}). No evidence for an eclipse indicates a low-to-mid
inclination of NSV 4838.

\subsubsection{superhump period change}

\begin{figure}
   \begin{center}
      \FigureFile(80mm,50mm){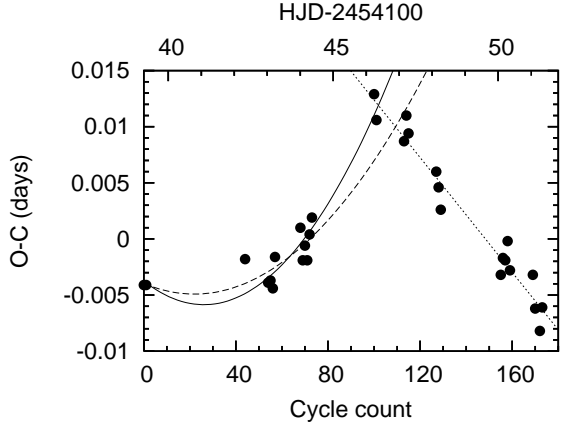}
   \end{center}
   \caption{$O-C$ diagram of the superhump maxima. The
 horizontal and vertical axes denote the cycle count and $O-C$,
 respectively. The solid and dashed curve correspond to the equation
 (2) and (3), respectively. After $E$=100, the data points can be fitted
 with the equation (4), which indicates that the superhump period
 unchanged in this phase.}\label{label5}
\end{figure}

We measured maximum timings of superhumps, which is summarized in table
\ref{table3}. A linear regression yields the following equation on the
superhump maximum timings:
\begin{equation}
HJD(max) = 2454139.2590(24) + 0.069802(21) \times E.
\end{equation}
Using this equation and table \ref{table3}, we obtained the $O - C$
diagram displayed in figure \ref{label5}. The solid curve in this figure
indicates the best fitting quadratic equation between $E$=0 and 101,
which is given by
\begin{eqnarray}
O - C =& - 3.75(1.11)\times10^{-3} - 1.62(0.41)\times10^{-4} E \nonumber \\
       & + 3.09(0.40)\times 10^{-6} E^{2}.
\end{eqnarray}
Also shown in this figure is the dashed curve by fitting a quadratic
equation between $E$=0 and 73. This is given by
\begin{eqnarray}
O - C =& - 3.97(1.05)\times10^{-3} - 8.57(6.66)\times10^{-4} E \nonumber \\
       & + 1.95(0.90)\times 10^{-6} E^{2}.
\end{eqnarray}
From these quadratic equations, we obtained
$P_{\rm dot} = \dot{P}/P = 8.9(1.1) \times 10^{\mathrm{-5}}$
from the former equation and
$\dot{P}/P = 5.6(2.6) \times 10^{\mathrm{-5}}$
from the latter equation, respectively. These results imply that the
superhump period increased at
$\dot{P}/P = 7(+3, -4) \times 10^{\mathrm{-5}}$ in the middle of the
plateau phase.

After $E=$100, the datapoints can be fitted with a linear equation. This
indicates that the superhump period remains almost constant during this
phase. A linear regression to the observed maximum timings between
$E$=100 and 174 yields
\begin{equation}
HJD(max) = 2454139.2969(20) + 0.06954(1) \times E.
\end{equation}
In conjunction with the $O - C$ diagram and the above equations, we
found that a sudden period decrease occurred between $E$=80 and
$E$=100.

 \begin{table}
 \begin{center}
 \caption{Timings of the superhump maxima.}
 \label{table3}
  \begin{tabular}{cc}
   \hline\hline
$E^*$ & HJD$^{\dagger}$ \\
   \hline
0 & 4139.2549 \\
1 & 4139.3247 \\
44 & 4142.3284 \\
54 & 4143.0243 \\
55 & 4143.0943 \\
56 & 4143.1634 \\
57 & 4143.2360 \\
68 & 4144.0064 \\
69 & 4144.0733 \\
70 & 4144.1444 \\
71 & 4144.2129 \\
72 & 4144.2850 \\
73 & 4144.3563 \\
100 & 4146.2519 \\
101 & 4146.3194 \\
113 & 4147.1551 \\
114 & 4147.2272 \\
115 & 4147.2954 \\
127 & 4148.1296 \\
128 & 4148.1980 \\
129 & 4148.2658 \\
155 & 4150.0748 \\
156 & 4150.1461 \\
157 & 4150.2157 \\
158 & 4150.2872 \\
159 & 4150.3544 \\
169 & 4151.0520 \\
170 & 4151.1188 \\
172 & 4151.2564 \\
173 & 4151.3283 \\
\hline
\multicolumn{2}{l}{$^*$ Cycle count.} \\
\multicolumn{2}{l}{$^{\dagger}$ HJD-2450000.} \\
\end{tabular}
\end{center}
\end{table}


\subsection{2005 June superoutburst}

\begin{figure}
   \begin{center}
      \FigureFile(80mm,50mm){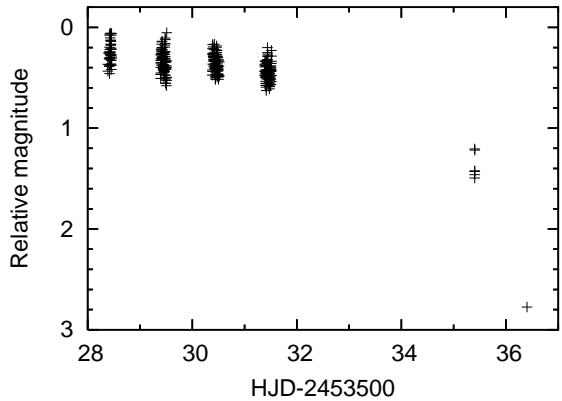}
   \end{center}
   \caption{Light curves of the 2005 June outburst obtained by
 JP. The horizontal and vertical axes indicate the HJD-2453000 and
 the relative magnitude, respectively.}\label{Figure6}
\end{figure}

\begin{figure}
   \begin{center}
      \FigureFile(80mm,50mm){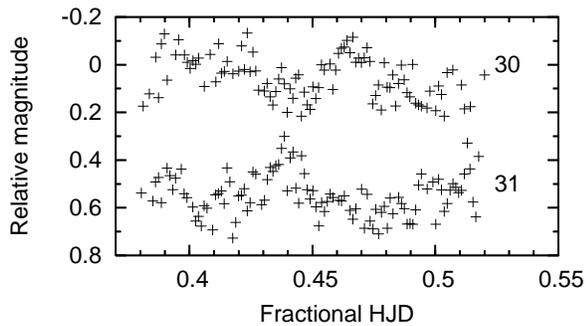}
   \end{center}
   \caption{Light curves of the 2005 June superoutburst obtained by
 JP.  The numbers in this figure indicate HJD-2453500. Although the data
 were noisy, superhumps are visible with an amplitude of 0.3
 mag.}\label{Figure7}
\end{figure}

The light curves during the 2005 June outburst are shown in figure
\ref{Figure6}. From HJD 2453528 to HJD 2453532, the magnitude declined at a
rate of 0.06(1) mag d$^{-1}$. Although this value is smaller compared to the
average decline rate of the plateau stage (0.10 mag d$^{-1}$,
\cite{war95book}), such a slow decline is sometimes observed at the late
phase of the plateau stage (\cite{bab00v1028cyg};
\cite{pat00dvuma}). Enlarged light curves are displayed in figure
\ref{Figure7}, in which hump-like features with an amplitude of 0.3 mag
are visible. After subtracting the linear declining trends, we also
performed the PDM for the data over the first 4 nights and found a weak
signal at 0.0699(1) days. This value is in good accordance with that
obtained during the 2007 February superoutburst. Judging from the above
results and other information such as the AAVSO light curve
generator\footnote{http://www.aavso.org/data/lcg/},
we conclude that we observed NSV 4838 from the late stage of the
superoutburst.

\subsection{mass ratio}

It is well known that the mass ratio of the system is related with the
fractional superhump excess (\cite{pat98evolution}; \cite{pat05suuma};
\cite{kni06secondary}). Here we use an empirical relation derived by
\citet{pat98evolution} as follows:  
\begin{equation}
\epsilon = \frac{0.23q}{1+0.27q},
\end{equation}
where $\epsilon$ is the fractional period excess, and $q$ is the mass
ratio of the system, respectively. Using $P_{\rm sh}$=0.0698 days and 
$P_{\rm orb}$=0.0678 days (Thorstensen, unpublished, see also
\cite{rit03cat}), we can estimate the mass ratio of NSV 4838 to be
0.13. This result is in good accordance with that using other
relations such as the equation (8) of \citet{pat05suuma}, equation (7)
of \citet{kni06secondary}.

\section{Discussion}

\subsection{long-term behavior of NSV 4838}

\begin{figure*}
\begin{center}
\resizebox{160mm}{!}{\includegraphics{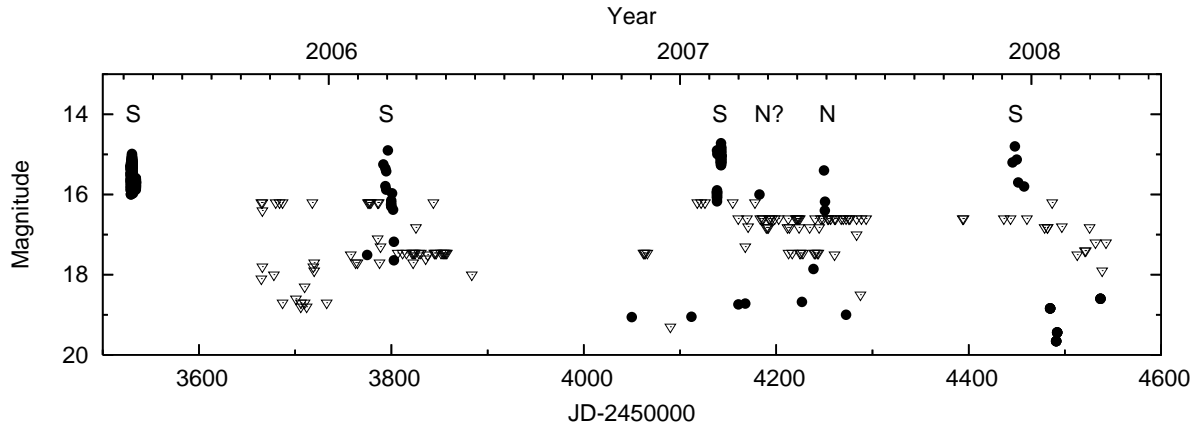}}
\end{center}
\caption{Long-term light curves of NSV 4838 since the 2005 June
 superoutburst using the AAVSO archive. The filled circles and bottom
 triangles indicate positive and negative observations,
 respectively. For display purpose, we precluded negative observations
 brighter than 16.0 mag. We also designate superoutburst and normal
 outburst as ``S'' and ``N'', respectively.}
\label{longlc}
\end{figure*}

The present photometric studies have shown the basic properties of
NSV 4838. The mean superhump periods, the fractional superhump excess,
and the duration of the plateau stage are quite a typical for the
"textbook" SU UMa-type dwarf novae \citep{war95book}. We also
investigated the long-term behavior of NSV 4838 based mainly on the
AAVSO light curve generator after the 2005
June superoutburst. The data include 229
negative observations and 104 positive observations from 2005 June to
2008 April. Although the short baseline of the archival data and its
faintness even at the maximum brightness cannot give a convincing result
as to whether some outbursts had been missed, we detected six outbursts
in total. Figure \ref{longlc} shows the long-term light curves of NSV 4838.
If we missed no superoutbursts, an estimated supercycle of NSV 4838 is
about 340 days, which is a typical value for SU UMa-type dwarf novae. On
the other hand, the total number of the recorded normal outburst is too
small for the estimated supercycle of NSV 4838 \citep{osa95wzsge}. One
possibility may be that the faintness of its maximum magnitude of normal
outburst prevented its detection. If we did not overlook any normal
outbursts, the event ratio of the normal outburst to superoutburst is
unusually small (see. e.g., table 1 of \citet{osa95wzsge}). Future
continuous observations of this object are needed to clarify the
activity of NSV 4838.

\subsection{distance}

According to an empirical relation of \citet{har04warner}, the
absolute magnitude of dwarf
novae at the maximum of the {\it normal outburst} is given as a function
of its orbital period,
\begin{equation}
M_{V} = 5.92 - 0.383 \times P_{\mathrm{orb}},
\end{equation}
where $P_{\rm orb}$ is in the unit of hour. According to multicolor photometry, the color index shows $V-R_{\rm c} \sim$ 0 near the bright maximum \citep{uem08j1021}. Thus the maximum magnitude in $V$ band may be almost the same value as that in $R_{\rm c}$ band. Since no filter is close to $R_{\rm c}$, it is not unreasonable that we regard the magnitude in non-filter as that in $V$ band. 

Extensive observations of
SU UMa-type dwarf novae suggest that the maximum magnitude of the normal
outburst is fainter than that of the superoutburst by about 0.5 mag
(\cite{war95book}; \cite{sha07vs0329}). In conjunction with the above facts including figure \ref{longlc}, a maximum magnitude of normal outburst for NSV 4838 is expected to be around 15.5 mag in $V$ band. Using this value, we roughly estimated 1.1 kpc as a distance to NSV 4838.

\subsection{superhump period change}

The general consensus concerning the superhump period change is that the
majority of SU UMa-type dwarf novae
show a decrease in their superhump periods during superoutburst
\citep{uem05tvcrv}. \citet{osa85SHexcess}
suggested that this is possibly due to shrinkage of the disk radius, or
consequence of mass depletion from the accretion disk. Recently, it
has been recognized that some SU UMa-type dwarf novae show an increasing
superhump period change during a plateau phase of superoutburst
(\cite{sem97swuma}; \cite{nog97alcom}). So far, we have confirmed this
increase for about 15 SU UMa-type dwarf novae, most of which have their
superhump periods shorter than 0.065 days \citep{oiz07v844her}.

Based on the refined thermal-tidal instability model proposed by
\citet{osa03DNoutburst}, the maximum radius during the superoutburst without
a precursor is larger than that with a precursor. When the accretion disk
reaches the tidal truncation radius (the maximum radius) at the onset of
the outburst, the matter will be piled up at this radius, at which the
stored matter works as a wall. The wall prevents the cooling wave from
propagating inward because the continuous tidal dissipation at the tidal
truncation radius urges matter to infall. In this case, no
precursor is observed at the onset of the superoutburst. On the other
hand, when the accretion disk does not reach the tidal truncation radius
at the onset of the outburst, the cooling wave can propagate from the
outermost annulus, which is observed as a rapid decline like a normal
outburst. If the eccentricity of the accretion disk grows to some
extent, the heating wave can again propagate from the outermost annulus,
because the tidal torque enhancement supplies the gas to the inner
annulus. This process is observed as a superoutburst with a precursor. 

Recently, \citet{uem05tvcrv} studied superhump period changes for the
2001 and 2004 superoutbursts of a short period SU UMa-type dwarf nova TV
Crv. It was found that the 2001 superoutburst showed no evidence of a
precursor and the superhump period increased during the plateau phase. On
the other hand, the 2004 superoutburst was accompanied by a precursor and
the superhump period kept almost constant during the plateau
phase. In combination with this observations and the refined
thermal-tidal instability model,
\citet{uem05tvcrv} suggested that enough propagation of the eccentric
mode outside the 3:1 resonance radius may be observed as an increase in
the superhump period. Therefore, short period SU UMa-type dwarf novae, in
which there exists a large annulus between the 3:1 resonance radius and
tidal truncation radius, tend to show a positive period derivative. On
the other hand, long period SU UMa-type dwarf novae and those with
a precursor during superoutburst, in which the gap
between the 3:1 resonance radius and the tidal truncation radius is
small, tend to show negative or zero period derivatives.

\begin{figure}
   \begin{center}
      \FigureFile(80mm,50mm){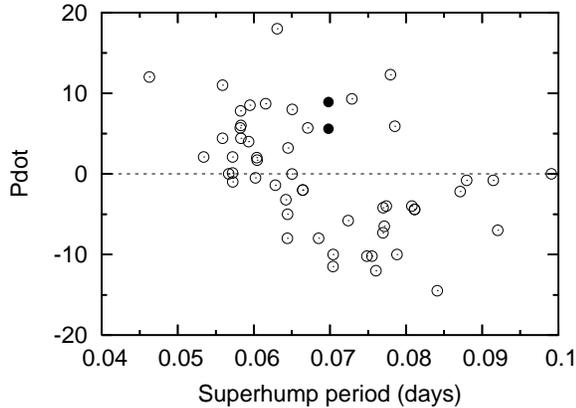}
   \end{center}
   \caption{$P_{\rm dot}$-$P_{\rm sh}$ diagram of SU UMa-type dwarf
 novae. The location of NSV 4838 in this figure is designated as filled
 circles. The original data were taken from \citet{oiz07v844her} (See also \citet{uem05tvcrv} for a comprehensive table of $P_{\rm dot}$ values.).}
\label{Figure9}
\end{figure}

Figure \ref{Figure9} illustrates the $P_{\rm dot}$-$P_{\rm sh}$ diagram
of SU UMa-type novae, in which we added the results of NSV 4838. It
should be noted that NSV 4838 showed the positive period derivatives
despite the fact that the superhump period of the system is as long as
0.070 days. In figure \ref{Figure9}, one can notice that a few
systems of $P_{\rm sh}>$0.07 days also show the positive $P_{\rm
dot}$. Such systems include VW CrB ($P_{\rm sh}$=0.07287 days,
\cite{nog04vwcrb}), TT Boo ($P_{\rm sh}$=0.07796 days,
\cite{ole04ttboo}), and RZ Leo ($P_{\rm sh}$=0.07853 days,
\cite{ish01rzleo}). Interestingly, these systems commonly show
long-lasting (exceeding 2 weeks), large-amplitude (exceeding 6 mag)
superoutursts and long supercycles. On the other hand, NSV 4838 exhibits relatively a small amplitude ($\sim$4 mag)\footnote{In dwarf novae during quiescence, $V$ magnitude is close to $g$ (e.g., \cite{szk05CVSDSS}). From this, $V$ magnitude in quiescence may be around 18.8, which corresponds to 3.7 in figure \ref{label1}.} and a moderate supercycle ($\sim$1 year). In addition, the negative $P_{\rm dot}$ was reported in the 1991 October superoutburst of EF Peg, which exhibited a long-lasting and large-amplitude superoutburst, as well as a long recurrence time (\cite{how93efpeg}; \cite{kat02efpeg}). Therefore, we suggest that these properties do not necessarily lead to the positive $P_{\rm dot}$.

Although the exact mechanism causing superhump period
changes still remains unknown, the present results imply that an
additional physical parameter on the accretion disk should be required
for explaining the observed diversity of the superhump period changes.
In order to clarify the nature of the superhump period changes, we should
collect further samples.

\section{Summary}

In this paper, we present CCD photometry of the dwarf nova NSV 4838
during the 2005 June and 2007 February superoutbursts, as well as the
long-term behavior of the system. During the 2007 February
superoutburst, the best estimated superhump period was 0.069824(83)d. We
also examined superhump period changes and concluded that the superhump
period increased with $\dot{P}$/$P$=+7(+3, -4)$\times 10^{\mathrm{-5}}$
. In the late phase of
the superoutburst, the mean superhump period became almost constant at
0.06954(1) days. Based on the long-term light curves, we found four
superoutbursts since 2005 June and estimated a supercycle of NSV 4838 to
be about 340 days. On the other hand, we detected at most two normal
outbursts. If we missed no normal outbursts, the event ratio of the normal
outburst to the superoutburst is unusually small for this supercycle. Using
the empirical relations, we derived about 1.1 kpc as the distance to NSV
4838 and 0.13 as the mass ratio, respectively. We plotted the obtained
values of NSV 4838 in the $P_{\rm dot}$-$P_{\rm sh}$ plane, finding that
the positions in the plane deviate from the general trend. In order to
understand the observed diversity of the superhump period changes,
additional physics should be invoked. This should be clarified in future
observations by collecting further information on superhump period changes.

\vskip 5mm

We acknowledge with thanks the variable star observations
from the AAVSO and VSNET International Database contributed by observers
worldwide and used in this research. This work is supported by a
Grant-in-Aid for the 21st
Century COE ``Center for Diversity and Universality in Physics'' from
the Ministry of Education, Culture, Sports, Science and Technology
(MEXT). This work is partly supported by a grant-in aid from
the Ministry of Education, Culture, Sports, Science and Technology
(No. 16340057, 17340055, 17740105, 18740153). Part of this work is
supported by a Research Fellowship of the Japan Society for the
Promotion of Science for Young Scientists (AI).

\end{document}